\documentclass[aps,prl,twocolumn,showpacs,preprintnumbers,amsmath,amssymb]{revtex4}
\usepackage{graphicx}% Include figure files
\usepackage{dcolumn}% Align table columns on decimal point
\usepackage{bm}% bold math

\newcommand {\beq} {\begin{equation}}
\newcommand {\eeq} {\end{equation}}
\newcommand {\beqa}{\begin{eqnarray}}
\newcommand {\eeqa}{\end{eqnarray}}
\newcommand {\del} {\partial}
\newcommand {\tr}{{\rm tr\,}}

\newcommand {\ee}{\mbox{e}}

\newcommand {\tphi}{\tilde{\phi}}

\begin{document}

%%%%%%%%%%%%%%%%%%%%%%%%%%%%%%%%%%%%%%%%%%%%%%%%%%%%%%%%%%%%%%%%%%%%
%  TITLE / AUTHOR                                                  %
%%%%%%%%%%%%%%%%%%%%%%%%%%%%%%%%%%%%%%%%%%%%%%%%%%%%%%%%%%%%%%%%%%%%

\title{
%Frequency-mode 
Non-lattice simulation for
supersymmetric gauge theories
in one dimension
%Proposal for non-lattice simulation of\\
%supersymmetric matrix quantum mechanics
}

\author{Masanori Hanada$^{1,2}$}
\email{hana@riken.jp}
%\cite{EmailKN}
\author{Jun Nishimura$^{3,4}$}
\email{jnishi@post.kek.jp}
\author{Shingo Takeuchi$^{4}$}
\email{shingo@post.kek.jp}
%\cite{EmailJN}

%\address{
\affiliation{
$^{1}$Department of Physics, Kyoto University,
%\\
Kitashirakawa, Kyoto 606-8502, Japan  \\
$^{2}$Theoretical Physics Laboratory, 
RIKEN Nishina Center,
%The Institute of Physical and Chemical Research (RIKEN),
%\\
2-1 Hirosawa, Wako, Saitama 351-0198, Japan \\
$^{3}$High Energy Accelerator Research Organization (KEK), 
		Tsukuba 305-0801, Japan \\
$^{4}$Department of Particle and Nuclear Physics,
School of High Energy Accelerator Science,
%\\
Graduate University for Advanced Studies (SOKENDAI),
Tsukuba 305-0801, Japan
%${}^3$Department of Particle and Nuclear Physics, 
%		The Graduate University for Advanced Studies (SOKENDAI),
%			Tsukuba, Ibaraki 305-0801, Japan
}

\date{June, 2007; preprint: RIKEN-TH-104, KEK-TH-1158
%, hep-th/yymmnnn
%\today %%new
}% It is always \today, today,

%%%%%%%%%%%%%%%%%%%%%%%%%%%%%%%%%%%%%%%%%%%%%%%%%%%%%%%%%%%%%%%%%%%%
%  ABSTRUCT    	                                                   %
%%%%%%%%%%%%%%%%%%%%%%%%%%%%%%%%%%%%%%%%%%%%%%%%%%%%%%%%%%%%%%%%%%%%

\begin{abstract}
Lattice simulation of supersymmetric gauge theories
is not straightforward.
%It is commonly considered
%that the lattice discretization is the only choice
%to study gauge theories on computer
%in a non-perturbative manner.
%% However, 
%% %due to the notorious problem with lattice fermions
%% %and the breaking of continuous translational symmetry,
%% its application to supersymmetric gauge theories
%% is not straightforward.
In some cases 
the lack of manifest supersymmetry just 
necessitates cumbersome fine-tuning, but 
in the worse cases the chiral and/or 
Majorana nature of fermions makes it difficult to 
even formulate an appropriate lattice theory.
We propose to circumvent all these problems
inherent in the lattice approach
by adopting a {\em non-lattice} approach
in the case of one-dimensional 
supersymmetric gauge theories,
which are important in the string/M theory context.
%both of these problems can be avoided
%by simulating the frequency modes
%using a {\em non-lattice} simulation
%in the case of one-dimensional 
%supersymmetric U($N$) gauge theories,
%In particular, our approach enables us to
%put M theory on computer.
%provides a perfect solution.
%Gauge symmetry is fixed by the static gauge, and the 
\end{abstract}


%11.25.-w; 11.25.Sq}
%11.25.-w Theory of fundamental strings
%11.25.Sq Nonperturbative techniques; string field theory

%%%%%
%11.10.Kk Field theories in dimensions other than four (see also 04.50.+h Gravity in more than four dimensions; 04.60.Kz Lower dimensional models in quantum gravity)
%11.15.-q Gauge field theories
%11.30.Pb Supersymmetry (see also 12.60.Jv Supersymmetric models)
%11.15.Ha Lattice gauge theory 
%11.15.Tk Other nonperturbative techniques
%%%%%%%%%

\maketitle

%%%%%%%%%%%%%%%%%%%%%%%%%%%%%%%%%%%%%%%%%%%%%%%%%%%%%%%%%%%%%%%%%%%%
%  1. INTRODUCTION                                                 %
%%%%%%%%%%%%%%%%%%%%%%%%%%%%%%%%%%%%%%%%%%%%%%%%%%%%%%%%%%%%%%%%%%%%

\paragraph*{Introduction.---}

%Since its invention, 
Lattice gauge theory,
together with the developments of
various simulation techniques,
has 
%been 
provided us with a powerful
non-perturbative method to study
%non-perturbative dynamics of 
gauge theories such as QCD.
% in particular.
However, when one tries to apply the method to
{\em supersymmetric} gauge theories, which
are interesting for many reasons, one has to
face with some practical and theoretical
obstacles. 

First of all, since the algebra
of supersymmetry contains continuous translations,
which are broken to discrete ones,
it seems unavoidable to break it
on the lattice.
%is difficult to keep it manifest.
%seems impossible to preserve it completely.
%unavoidable to preserve it completely.
%keep the full supersymmetry.
Then, one has to include all the
relevant terms allowed by symmetries preserved
on the lattice, and fine-tune the coupling constants
to arrive at the desired supersymmetric fixed point
in the continuum limit.
Recent progress (See ref.\ \cite{Giedt:2007hz} and
references therein.)
%for a review.)
%in this direction 
is that
%one can achieve this by
%only a few parameters (or no parameters, in some case)
%(or no) fine-
one can reduce the number of parameters to
be fine-tuned (even to zero
%, i.e., no fine-tuning,
in some cases)
%, or even do without fine-tuning,
by preserving some part of supersymmetry. 
%When the space-time dimensionality 
In lower dimensions, one can alternatively
take the advantage of super-renormalizability, 
and determine the appropriate counter-terms 
by perturbative calculations
to avoid fine-tuning.
%thereby avoiding fine-tuning.
%so that fine-tuning is no more necessary.
These two approaches can be nicely illustrated
in one dimension by the example of 
a supersymmetric 
anharmonic oscillator 
\cite{Catterall:2000rv,Giedt:2004vb}.
%, which we discuss later.
% ``exact'' 
%results can be obtained by solving 
%the Schr\"odinger equation.

In the string/M theory context,
supersymmetric gauge theories appear
in various ways. In particular,
the low-energy effective theory
of a stack of $N$ $p$-branes are given by
$(p+1)$-dimensional U($N$) super Yang-Mills theory.
This led to various interesting conjectures.
For instance,
%By now there are many evidences for
the gauge/gravity duality states
that the strong coupling limit of 
large-$N$ gauge theories has a dual description
in terms of classical supergravity.
A different but closely related conjecture
asserts that non-perturbative formulations
of superstring/M theory can be given
by matrix models, which can be obtained
by dimensionally reducing 10d ${\cal N}=1$ 
U($N$) super Yang-Mills theory to $D=0,1,2$ dimensions.
In particular, the $D=1$ model \cite{BFSS} corresponds
to the M Theory.
Its bosonic version
has been studied by
Monte Carlo simulation
% of bosonic theories
%(omitting fermions by hand) performed 
%in the ``quenched approximation''
%omitting  by hand
using the lattice formulation 
\cite{latticeBFSS} and the
continuum quenched Eguchi-Kawai model
\cite{Kawahara:2005an}.
However, the next step of including
fermionic matrices would be difficult
even theoretically because of 
%transform as
their Majorana-Weyl nature inherited from 10d.
%nature of fermions.
%of the chiral and/or 
%Majorana nature of fermions.
See ref.\ \cite{Kaplan:2005ta} for a proposal
using lattice, 
which preserves half of SUSY at the expense
of breaking the SO(9) symmetry.
%invariance.
%In the lattice approach, in particular,
%it has not been possible to even write down 
%an SO(9) invariant action \cite{Kaplan:2005ta}.
% due to the notorious problem of lattice fermions.

The aim of this letter
%the present article 
is to
point out that
% in one dimension, 
there exists
an extremely simple and elegant method
to simulate supersymmetric gauge theories
in one dimension. 
(We discuss the case
of U($N$) gauge group, but extension to 
more general group must be possible.)
%propose a non-lattice simulation method for 
%non-perturbative studies of 
%supersymmetric gauge theories
%in one dimension. 
In particular,
it will enable us to study the
gauge theory side of
the aforementioned duality
%from first principles, 
and also to put M theory on computer.
%in an extremely simple and elegant manner.

Let us first recall that
the importance of the lattice formulation lies
in its manifest gauge invariance.
% it is the manifest gauge invariance
%that makes the lattice formulation
%the unique choice for a non-perturbative
%formulation in ordinary gauge theories.
In the present 1d case, however,
%may be viewed as 1d gauge theory,
%In the present case of 1d gauge theory, however,
the gauge dynamics is almost trivial.
(We assume that the 1d direction
is compact.
% having in mind
%a setup for 
%% an application to 
%finite temperature.
% and the 
%Eguchi-Kawai reduction \cite{EK}.
%
%, for instance. 
The non-compact case would be easier since
the gauge dynamics is completely trivial.)
%If one has free boundary conditions in 1d,
%the gauge field can be gauged away.
%If the 1d direction is compact, 
%some global degrees of freedom remain
%due to nontrivial holonomy.
This gives us an opportunity to use a non-lattice 
formulation. 
More specifically,
we first take the static diagonal gauge.
Using the residual large gauge transformation,
we can choose a gauge slice such that
the diagonal elements of the constant gauge field 
lie within a minimum interval.
Finally we expand the fields into Fourier
modes, and keep only the modes 
up to some cutoff.
% $\Lambda$. 
The crucial point of our method is that
%By construction, 
the gauge symmetry is completely
fixed 
(up to the global permutation group, which is
kept intact)
%of U($N$) indices)
%global SU($N$) 
before introducing the cutoff.
This is specific to one dimension,
% possible only in one dimension.
and the momentum cutoff regularization 
in higher dimensions generally breaks gauge invariance.

\paragraph*{Supersymmetric anharmonic oscillator.---}

To gain some insight into 
our new
%the present
approach, we first apply it to 
a non-gauge supersymmetric theory,
%the system of a supersymmetric anharmonic oscillator,
which is well studied by the lattice formulation.
In particular, 
%we will see that 
supersymmetry,
which is broken by the cutoff in our formalism,
is restored much faster than the continuum limit
is achieved.

\begin{figure}[htb]
\begin{center}
\includegraphics[height=6cm]{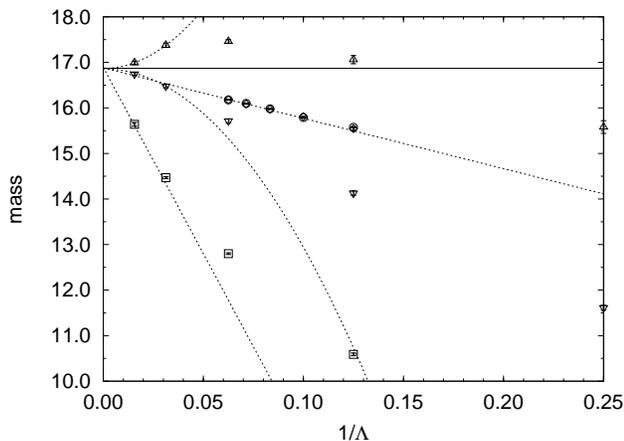}
%{bfss_continuum.eps}
\end{center}
\caption{The circles (diamonds)
%(rhombi)
are the mass for the boson (fermion)
obtained by our method for $\Lambda =8,10,12,14,16$.
% are plotted against $1/\Lambda$.
%The results obtained from lattice formulations
%are also plotted with the identification $\Lambda = L/2$,
%where $L$ is the number of sites.
The triangles (inverted triangles) 
are the mass for the boson (fermion)
obtained by Giedt et al.\ \cite{Giedt:2007hz}
with the O($a$)-improved lattice action,
and the squares are the results obtained by Catterall and
Gregory \cite{Catterall:2000rv}
with the lattice action preserving half of SUSY,
hence degenerate. The horizontal line represents
the exact result,
% (16.87) obtained by solving the Schr\"odinger equation, 
and the dotted lines
represent the expected behaviors at large $\Lambda$.
}
\label{anh_oscil_compare}
\end{figure}

While the manuscript was being prepared,
we received a preprint \cite{Bergner:2007pu},
in which the same model is studied 
on the lattice using various methods.
As far as non-gauge theories are concerned,
our approach is almost equivalent to 
the method with
the non-local SLAC derivative \cite{SLAC}.
The only difference is
the identification of the
modes at the boundary of the Brillouin zone
in the lattice case.
As a consequence, our results shown in
fig.\ \ref{anh_oscil_compare} agree
with the corresponding 
results in ref.\ \cite{Bergner:2007pu}.

%intensively.
The model
%action of the 
%supersymmetric 
%model
%quantum mechanical system reads
is defined by the action
%given by
\beq
S=\int_0 ^{\beta} dt
\left[ \frac{1}{2} \left\{ 
%\left(\frac{d \phi }{d t}
( \del_t \phi )^2
%\dot{\phi}^2 
+ {h'}(\phi)^2  \right\}
+ \bar{\psi} ( \del_t + {h''}(\phi) ) \psi \right] \ ,
\label{action}
\eeq
where $\phi$ is a real scalar field,
and $\psi$ is a one-component Dirac field,
both in 1d, obeying periodic boundary conditions.
This model is supersymmetric for arbitrary 
function $h(\phi)$, but here we take
$h(\phi)=\frac{1}{2} m \phi ^2 + \frac{1}{4} g \phi ^4$.
%with positive $g$ following refs.
%The symbol $'$ in (\ref{action}) represents
%the derivative with respect to 
%
%The coupling constant $g$ is assumed to be positive,
%in which case the supersymmetry is not spontaneously broken.
In our approach we make a Fourier expansion 
\beq
\phi (t) = \sum_{n=-\Lambda}^{\Lambda} 
\tilde{\phi}_n \ee^{i \omega n t} \ ; \quad
\omega\equiv\frac{2\pi}{\beta}
\eeq
and similarly for the fermionic fields,
where $n$ takes integer values,
and $\Lambda$ is the UV cutoff.
% \beqa
% \phi (t) &=& \sum_{n=-\Lambda}^{\Lambda} \phi_n \ee^{i \omega n t} \\
% \psi (t) &=& \sum_{n=-\Lambda}^{\Lambda} \psi_n \ee^{i \omega n t} \\
% \bar{\psi} (t) &=& \sum_{n=-\Lambda}^{\Lambda}
% \bar{\psi}_n \ee^{- i \omega n t} \ ,
% \eeqa
In terms of the Fourier modes, the action
can be written as $S=S_{\rm B}+S_{\rm F}$, where
\beqa
S_{\rm B} &=&   \beta
\left[ \,
 \sum_{n=-\Lambda}^{\Lambda} 
  \frac{1}{2} \Bigl\{ 
 (n \omega)^2 + m^2 \Bigr\} 
\tphi_n \tphi_{-n}  \right. \nonumber \\ 
%+  m g (\tphi * \tphi * \tphi)_{n} \tphi_{-n}
&~& \left.
+  m g  (\tphi^4)_{0}
% \tphi_{-n}
 + \frac{1}{2} 
%g^2 (\tphi * \tphi *\tphi * \tphi * \tphi)_{n} \tphi_{-n}
g^2 (\tphi ^6)_{0}
%(\tphi ^5)_{n} \tphi_{-n}
\right] \ ,
% \nonumber 
\\
%\beta \left[ \sum_{n=-\Lambda}^{\Lambda} 
S_{\rm F} 
%&=& 
%\beta \sum_{n=-\Lambda}^{\Lambda} 
%\Bigl[ (i n \omega  + m) \tilde{\bar{\psi}}_n \tilde{\psi}_n
%+ 3g \tilde{\bar{\psi}}_n (\tphi * \tphi * \tilde{\psi})_n
%\Bigr] \nonumber \\
&=& \sum_{nk}
\tilde{\bar{\psi}}_n {\cal M}_{nk} \tilde{\psi}_k \ ,
\nonumber \\
{\cal M}_{nk}
&=& \beta \Bigl[
(i n \omega  + m) \delta_{nk}
%+ 3 g (\tphi * \tphi )_l \delta_{n,k+l}
+ 3 g (\tphi ^2)_l \delta_{n,k+l}
\Bigr] \ .
\label{action_Fourier}
\eeqa
We have introduced a shorthand
notation
%the generalized convolution
%the symbol 
\beq
\Bigl(f^{(1)}  \cdots  f^{(p)}\Bigr)_n 
\equiv \sum_{k_1 + \cdots + k_{p}=n}
f^{(1)}_{k_1} \cdots f^{(p)}_{k_p} \ .
\eeq
%represents 
Integrating out the fermions first, we obtain
the effective action for the bosons as
\beq
S_{\rm eff} = S_{\rm B}
- \ln \det {\cal M} \ ,
\label{eff-action}
\eeq
where the fermion determinant $\det {\cal M}$ is
real positive for positive $m$ and $g$.

As an efficient algorithm to simulate the model
(\ref{eff-action}), we use the idea of 
the hybrid Monte Carlo algorithm \cite{Duane:1987de}.
Let us introduce the auxiliary real field
$\Pi(t)$, whose Fourier components are denoted
as $\tilde{\Pi}_n$, and consider the action
\beq
S_{\rm HMC} 
= S_{\rm eff} + \sum_{n=-\Lambda}^{\Lambda} 
  \frac{1}{2} \tilde{\Pi}_n \tilde{\Pi}_{-n}  \ .
\label{action_HMC}
\eeq
Integrating out the auxiliary field first,
we retrieve (\ref{eff-action}).
In order to update the fields,
we solve the 
%evolution 
equations 
\beqa
\frac{d \tphi_n(\tau)}{d \tau}
&=& \alpha_n \frac{\del S_{\rm HMC}}{\del \tilde{\Pi}_n} 
= \alpha_n \tilde{\Pi}_{-n} 
\label{phi_evolution}
\\
\frac{d \tilde{\Pi}_n(\tau)}{d \tau}
&=& - \alpha_n \frac{\del S_{\rm HMC}}{\del \tphi_n} 
= - \alpha_n \frac{\del S_{\rm eff}}{\del \tphi_n}  
\label{pi_evolution}
\eeqa
along the fictitious time $\tau$ for 
a fixed interval
% $\Delta$.
$\tau_{\rm f}$.
The real coefficients $\alpha_n$ should be
optimized based on the idea of
the Fourier acceleration \cite{Catterall:2001jg}.
%as well as $\tau_{\rm f}$ %$\Delta$
%and the step size $\Delta\tau$
%are the parameters that should be 
%optimized \cite{Catterall:2001jg}.)
This evolution, if treated exactly, 
conserves the action $S_{\rm HMC}$.
In practice, we discretize the
$\tau$-evolution in such a way
(the leap-frog discretization)
that the reversibility is maintained. 
%particular
%fictitious time evolution in a particular
%way 
%which maintains the reversibility.
Due to the discretization, the action
$S_{\rm HMC}$ changes by a small amount,
say $\Delta S_{\rm HMC}$.
In order to satisfy the detailed balance,
we accept the new configuration with 
the probability 
$\min (1, \ee^{- \Delta S_{\rm HMC}})$,
which is the usual Metropolis procedure.
Before we start a new $\tau$-evolution,
we refresh the $\tilde{\Pi}_n$ variables
by drawing Gaussian random numbers
%distribution 
which follow from the action (\ref{action_HMC}).
This procedure is necessary for avoiding
the ergodicity problem.
(The step size $\Delta \tau$ should
be optimized for fixed $\tau_{\rm f}$
by maximizing
%in such a way that
the acceptance rate times $\Delta \tau$.
%becomes maximum.
%should be kept reasohigh enough
%The acceptance rate should be kept reasohigh enough
%higher than 70\% 
%by making 
%the discretization of $\tau$ fine 
%small enough.
Then $\tau_{\rm f}$ should be optimized by
minimizing the autocorrelation time 
in units of step in the $\tau$-evolution.)

The main part of the computation is
the evaluation of the 
term in eq.\ (\ref{pi_evolution})
given by
\beqa
  \frac{\del S_{\rm eff}}{\del \tphi_n}  
&=& 
  \beta
\Bigl[ \{  (n\omega)^2 + m^2 \} \tphi_{-n}
 + 4 mg (\tphi^3)_{-n} \\
&~& + 3 g^2 (\tphi^5)_{-n} \Bigr] 
 -  \tr \left(\frac{\del {\cal M}}{\del \tphi_n} 
{\cal M}^{-1}\right) \ .
\eeqa
The convolution requires O($\Lambda^2$) 
calculations, while 
%the calculation of 
the inverse ${\cal M}^{-1}$ requires
O($\Lambda^3$) calculations.

As usual, we extract masses
from the exponential decay of the two-point functions
\beqa
G_{\rm B}(t)
&\equiv& 
\langle \phi(0) \phi(t) \rangle
= b_0
+ 2 \sum_{n=1}^{\Lambda}
b_n
\cos(\omega n t)  \ , 
\label{prop_b} \\
G_{\rm F}(t)
&\equiv& 
\langle \psi(0) \bar{\psi}(t) \rangle
= \sum_{n=-\Lambda}^{\Lambda} 
c_n \, \ee^{-i\omega n t} \ ,
\label{prop_f}
\eeqa
%% \beqa
%% G_{\rm 1B}(t)
%% &=& \frac{1}{\beta} \int_{0}^{\beta} ds \, 
%% \langle \phi(s) \phi(s+t) \rangle \nonumber \\
%% &=& b_0
%% + 2 \sum_{n=1}^{\Lambda}
%% b_n
%% \cos(\omega n t)  \ , 
%% \label{prop_b} \\
%% G_{\rm 1F}(t)
%% &=& \frac{1}{\beta} \int_{0}^{\beta} ds \, 
%% \langle \psi(s) \bar{\psi}(s+t) \rangle \nonumber \\
%% &=& \sum_{n=-\Lambda}^{\Lambda} 
%% c_n \, \ee^{-i\omega n t} \ ,
%% \label{prop_f}
%% \eeqa
where we have defined 
$b_n \equiv \langle | \tphi_n |^2 \rangle$ and
%$\langle (\tphi_0)^2 \rangle$
$c_n \equiv \langle ( {\cal M}^{-1})_{nn}\rangle $. 
%In the continuum limit the fermion 2-point function
%is expected to $G_{\rm 1F}(t)=0$ for $t<0$.
%In order to extract the fermion mass,
For the fermion,
it proved convenient to consider ,
instead of (\ref{prop_f}), a symmetrized correlator
\beqa
G_{\rm F}^{\rm (sym)}(t)
&\equiv& \frac{1}{2} \{ G_{\rm F}(t) + G_{\rm F}(-t) \} 
\nonumber \\
&=& c_0
+ 2 \sum_{n=1}^{\Lambda} {\rm Re} (c_n) \cos (\omega n t) \ ,
\label{prop_f_sym}
\eeqa
where we have used the fact
$({\cal M}_{nk})^*={\cal M}_{-n, -k}$.
In fact
%It turns out that 
the functions
(\ref{prop_b}) and (\ref{prop_f_sym})
with respect to $t$ oscillate
with the frequency of the order of cutoff.
This is nothing but the Gibbs phenomenon
due to 
%In the present formulation , however, 
the sharp cutoff in the sum over Fourier modes.
%on in 
%(\ref{prop_b}) and (\ref{prop_f_sym}) makes 
%the corresponding functions of $t$ oscillate
%with the frequency of the order of cutoff.
%O($\omega \Lambda$).
To overcome this problem, we note that
%we exploit the fact that
the coefficients $b_n$ behave as
$b_n \sim \frac{d_1}{n^2} + \frac{d_2}{n^4}$
%% \beq
%% b_n \sim \frac{d_1}{n^2} + \frac{d_2}{n^4} 
%% \label{asymptotic}
%% \eeq
at large $n$ 
as can be shown from general arguments.
We obtain the coefficients $d_1$ 
and $d_2$ from the results at
$n=\Lambda-1,\Lambda$, and extend the sum
in (\ref{prop_b}) over $n$ up to 1000
assuming the above asymptotic form.
% (\ref{asymptotic}).
%The function of $t$ obtained in this way
%are shown in fig.\ 
We make an analogous analysis for
${\rm Re}(c_n)$ in (\ref{prop_f_sym}).
In this way we are able to see clear exponential behaviors,
and extract the corresponding masses.
The results for $\beta=1$, $m=10$, $g=100$
are plotted against $1/\Lambda$
in fig.\ \ref{anh_oscil_compare}.
(Note that the effective coupling constant is
$g/m^2=1$.)
%, which is O(1), and the finite $\beta$ effects
%are negligible.)
%
%The mass for the boson and fermion
%, which we denote as $m_{\rm B}$
%and $m_{\rm F}$, respectively, 
We find that the finite $\Lambda$ effects
are O($1/\Lambda$),
%they behave as $m = m_0 + c/\Lambda$
and that 
the data points 
for the boson and the fermion
lie on top of each other.
%
%the masses for the boson and the fermion
%$m_{\rm B}$ and $m_{\rm F}$ 
%agree very well.
%(The data points lie on top of each other.)
Thus in our formalism,
the effect of supersymmetry 
breaking by the cutoff
disappears much faster than $1/\Lambda$.
%although the supersymmetry is 
%violated at the cutoff scale, its effect seems to
%disappear much faster than $1/\Lambda$.
In the same figure we also plot the results obtained from
lattice formulations for comparison.
(Matching the number of degrees of freedom,
we make an identification $\Lambda=\frac{L}{2}$,
where $L$ is the number of sites.)

%%%%% BFSS_continuum. %%%%%
%%%%%%%%%%%%%%%%%%

\paragraph*{Supersymmetric matrix quantum mechanics.---}
%\paragraph*{1d SUSY gauge theory.---}

%Let us study the supersymmetric matrix quantum mechanics.
%which has been studied
%in refs.\ by the standard lattice formulation.
Here we consider a model with four supercharges
defined by
%by the action
%The action is given by
\beqa
S
%_{\rm mQM} 
&=& 
\frac{1}{g^2} \int_0^{\beta}  
% _{-\infty} ^{\infty} 
d t \, 
\tr \left\{ 
\frac{1}{2} (D_t X_i)^2 - 
\frac{1}{4} [X_i , X_j]^2  
\right. \nonumber \\
&~& 
%\left.
+ \bar{\psi} D_t \psi 
- \bar{\psi} \sigma_i [X_i , \psi ]
\Bigr\} \ ,
\label{cQM}
\eeqa
where $D_t  = \del_t
%\frac{\partial}{\partial t} 
  - i \, [A(t), \ \cdot \ ]$ represents the covariant derivative
with the gauge field $A(t)$ being an $N\times N$ Hermitian matrix.
%iti1d gauge field.
The bosonic matrices $X_i(t)$  $(i=1,2,3)$
are $N\times N$ Hermitian,
and the fermionic matrices $\psi_\alpha(t)$ and 
$\bar{\psi}_\alpha(t)$  $(\alpha=1,2)$
are $N\times N$ matrices with complex Grassmann entries.
The $2\times 2$ matrices $\sigma_i$ 
%in (\ref{cQM}) 
are the Pauli matrices.
The model can be obtained formally by dimensionally
reducing 4d ${\cal N}=1$ U($N$) super
Yang-Mills theory to 1d,
and it can be viewed as 
a one-dimensional U($N$) gauge theory.
(The totally reduced model has been studied
by Monte Carlo simulation in refs.\ \cite{4dsusy}.)
Let us assume the boundary conditions
to be periodic for bosons
and anti-periodic for fermions.
The extent $\beta$ in the Euclidean time 
direction then corresponds to the inverse
temperature $\beta \equiv 1/T$.
The parameter $g$ in (\ref{cQM})
can always be absorbed
%scaled out 
by an appropriate rescaling of the matrices and 
the time coordinate $t$.
Hence we set $g = \frac{1}{\sqrt{N}}$ without loss 
of generality.

Let us take the static diagonal gauge
$A(t) = \frac{1}{\beta} {\rm diag}
(\alpha_1 , \cdots , \alpha_N)$,
%% \beq
%% A(t) = \frac{1}{\beta} {\rm diag} 
%% (\alpha_1 , \cdots \alpha_N) \ ,
%% \eeq
where $\alpha_a$ ($a = 1 , \cdots , N$) can be chosen to 
lie within the interval $(-\pi , \pi]$
%in the range $ \alpha_i \in (-\pi , \pi]$
by making a gauge transformation
with a non-zero winding number \cite{endnote}.
% \cite{endnote2}.
% Then the covariant derivative in (\ref{cQM}) is replaced by
% \beq
% (D_t X_\mu)_{ij}
% = \del_t X_\mu^{ij} - 
% i \frac{\alpha_i - \alpha_j}{\beta} 
% X_\mu^{ij} \ ,
% \eeq
% and
We have to add to the action a term
\beq
S_{\rm FP} =
- \sum_{a<b} 2 \ln 
\left| \sin \frac{\alpha_a - \alpha_b}{2}
\right|  \ , 
\eeq
which appears from the Faddeev-Popov procedure,
and the integration measure for $\alpha_a$
is taken to be uniform.

We make a Fourier expansion 
\beq
X_i ^{ab} (t) = \sum_{n=-\Lambda}^{\Lambda} 
\tilde{X}_{i n}^{ab} \ee^{i \omega n t} \ ; \
\psi_\alpha ^{ab} (t) = \sum_{r=-\lambda}^{\lambda}
\tilde{\psi}_{\alpha r}^{ab} \ee^{i \omega r t} \ ,
\eeq
%% \beqa
%% X_\mu ^{ij} (t) &=& \sum_{n=-\Lambda}^{\Lambda} 
%% \tilde{X}_{\mu n}^{ij} \ee^{i \omega n t} \ , \\
%% \psi_\alpha ^{ij} (t) &=& \sum_{r=-\lambda}^{\lambda}
%% \tilde{\psi}_{\alpha r}^{ij} \ee^{i \omega r t} \ ,
%% \eeqa
and similarly for $\bar{\psi}$,
where $r$ takes half-integer values, 
due to
%taking account of 
the anti-periodic boundary conditions,
and $\lambda \equiv \Lambda-1/2$.
%Here we have introduced the UV cutoff $\Lambda$.
Eq.\ (\ref{cQM}) can then be written as
\beqa
S &=&  N \beta
\left[
\frac{1}{2} \sum_{n=-\Lambda}^{\Lambda} 
%\left\{ n \omega - \frac{1}{\beta} (\alpha_i - \alpha_j)
\left\{ n \omega - \frac{\alpha_a - \alpha_b}{\beta} 
\right\} ^2   \tilde{X}_{i , -n}^{ba} \tilde{X}_{i n}^{ab}
\right. \nonumber \\ 
&~& 
\left. 
- \frac{1}{4} 
 \tr \Bigl( [ \tilde{X}_{i} , \tilde{X}_{j}]^2  \Bigr)_0
%- \Bigl( \tr (\tilde{X}_{\mu} \tilde{X}_{\nu})^2 \Bigr)_0 
\right] 
\nonumber \\
&~& + N \beta \sum_{r=-\lambda}^{\lambda} \Biggl[
i \left\{ r \omega - \frac{\alpha_a - \alpha_b}{\beta} 
\right\} 
\tilde{\bar{\psi}}_{\alpha r}^{ba} \tilde{\psi}_{\alpha r}^{ab} 
 \nonumber \\
&~& - (\sigma_i)_{\alpha\beta}
 \tr \Bigl\{ \tilde{\bar{\psi}}_{\alpha r} \Bigl(
[ \tilde{X}_{i},\tilde{\psi}_{\beta}] \Bigr)_r \Bigr\} \Biggr] \ .
\label{bfss_action_cutoff}
\eeqa
%As a residual 
%While the static diagonal gauge fixes the local gauge
%symmetry completely, 
%
% The above gauge-fixed action has a symmetry under
% the large gauge transformation
% \beqa
% \alpha_i &\mapsto& \alpha_i + 2 \pi \nu_i \\
% \tilde{X}_{\mu n}^{ij}
% &\mapsto&
% \tilde{X}_{\mu n+\nu_i-\nu_j}^{ij} \ ,
% \eeqa
% were it not for the UV cutoff.

\begin{figure}[htb]
\begin{center}
\includegraphics[height=6cm]{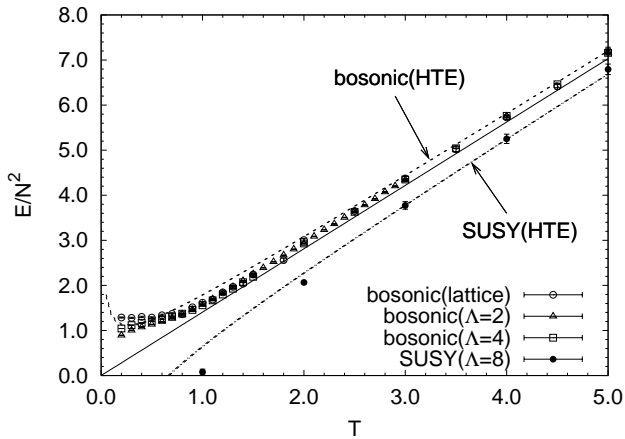}
%{bfss_continuum.eps}
\end{center}
\caption{
The energy (normalized by $N^2$)
is plotted against 
%versus 
temperature for the matrix quantum mechanics
%i1d supersymmetric gauge theory 
with $N=4$. 
%Both bosonic and SUSY cases are shown.
}
\label{energy}
\end{figure}

\begin{figure}[htb]
\begin{center}
\includegraphics[height=6cm]{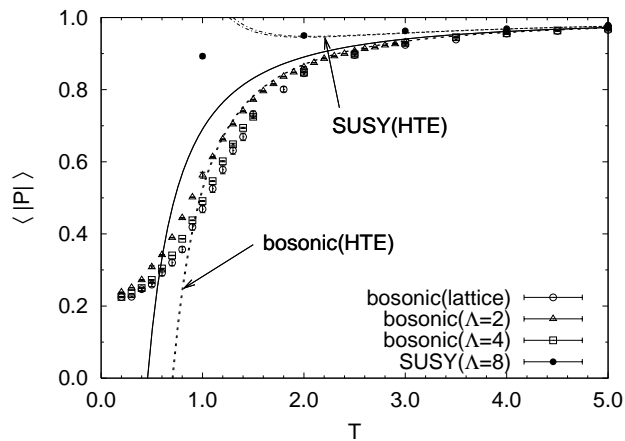}
%{bfss_continuum.eps}
\end{center}
\caption{
The same as fig.\ \ref{energy} 
%except that here 
but for the absolute value of 
the Polyakov line.
%The Polyakov line versus temperature for $N=4$.
}
\label{polyakov}
\end{figure}

The algorithm for simulating (\ref{bfss_action_cutoff})
is analogous to the previous model.
Here we introduce the auxiliary variables
$\Pi_{i}(t)$ and $p_a$, which are $N\times N$
Hermitian matrices conjugate to $X_i(t)$
and $N$ real variables conjugate to $\alpha_a$,
respectively.
The fermion determinant is real positive,
and the computational effort for one step in 
%solving 
the $\tau$-evolution 
%equations 
is proportional to $\Lambda^3 N^6$.
%of the order of
%O($\Lambda^3 N^6$).
%
%The total action then takes the form
%\beq
%S_{\rm}
%\eeq
%Using the new approach, we study the $N=4$ case. 
Figures \ref{energy} and 
\ref{polyakov} show the results for 
the energy and the Polyakov line, respectively,
for the bosonic and supersymmetric cases.

%%%%% BFSS_continuum. %%%%%
%%%%%%%%%%%%%%%%%%

%%%%% BFSS_continuum. %%%%%
%%%%%%%%%%%%%%%%%%

In the bosonic case we also plot the results
from lattice simulation with the lattice spacing
$a=0.02$.
%, which is small enough \cite{Kawahara:2005an}.
(The number of lattice sites
is given by $L=1/(Ta)$, which is 50 for $T=1$.)
% and 100 at $T=0.2$.)
The results obtained by our new method
approach the lattice result
as $\Lambda$ is increased. 
%In fact even $\Lambda=2$ 

In the supersymmetric case, our preliminary
results with $\Lambda=8$ reproduce the 
asymptotic behavior at large $T$
obtained by the high temperature 
expansion (HTE) \cite{HTE}
up to the next-leading order.
%but we start to see some deviation at $T=2$.
(The solid lines represent the results
at the leading order of HTE, which are the same for
the bosonic and SUSY cases.)
Note that our method is applicable also 
at low temperature, where the HTE is no more valid.
%The results shall be reported elsewhere.
%We report on the results 
%shall be 
%but it requires more effort
%due to the necessity to increase $\Lambda$
%continues to be applicable
%in the low temperature regime.
%which is interesting
%from the viewpoint of the gauge/gravity duality.
%
%The results shall be reported in a forth-coming paper.

%%%%%%%%%%%%%%%%%%%%%%%%%%%%%%%%%%%%%%%%%%%%%%%%%%%%%%%%%%%%%%%%%%%%
%  7. SUMMARY                                                      %
%%%%%%%%%%%%%%%%%%%%%%%%%%%%%%%%%%%%%%%%%%%%%%%%%%%%%%%%%%%%%%%%%%%%

\paragraph*{Summary and concluding remarks.---}
In this letter we have proposed a new simulation method,
which enables non-perturbative studies
of supersymmetric gauge theories in one dimension.
For practical implementation, the idea of 
the hybrid Monte Carlo algorithm seems to be quite useful.
In particular, the Fourier acceleration
requires no extra cost, since 
we deal with the Fourier modes directly.
The continuum limit is achieved much faster than
one would expect naively from the number of degrees of freedom.
This is understandable since 
%the Fourier modes diagonalize
%the kinetic term of the action, and therefore 
the Fourier modes omitted
by the cutoff scheme are 
%the ones that are 
naturally suppressed by the kinetic term.
%We emphasize that the results for the supersymmetric gauge
%theory presented in this letter would have been
%difficult to obtain by the conventional lattice approach.

It is straightforward to apply our method to 
the most interesting case with sixteen supercharges,
which is currently under investigation \cite{Mtheory}.
%Considering the computational effort of the order of
%O($\Lambda ^3 N^6$), 
%We hope to report on that case in the future publications.

%%%%%%%%%%%%%%%%%%%%%%%%%%%%%%%%%%%%%%%%%%%%%%%%%%%%%%%%%%%%%%%%%%%%
%  ACKNOWLEDGEMENTS                                                %
%%%%%%%%%%%%%%%%%%%%%%%%%%%%%%%%%%%%%%%%%%%%%%%%%%%%%%%%%%%%%%%%%%%%

\paragraph*{Acknowledgements.---}
The authors would like to thank S.\ Catterall,
Y.\ Kikukawa and F.\ Sugino
for valuable discussions.
The simulation has been performed
% workstations at
using Yukawa Institute Computer Facility,
RIKEN Super Combined Cluster System
and a PC cluster at KEK.

%%%%%%%%%%%%%%%%%%%%%%%%%%%%%%%%%%%%%%%%%%%%%%%%%%%%%%%%%%%%%%%%%%%%
%  REFFERENCE                                                      %
%%%%%%%%%%%%%%%%%%%%%%%%%%%%%%%%%%%%%%%%%%%%%%%%%%%%%%%%%%%%%%%%%%%%

%\end{references}


\begin{thebibliography}{99}
\bibitem{endnote}
In the present case with only adjoint representations,
it is more convenient to impose
a weaker constraint
$\max_i(\alpha_i) - \min_i(\alpha_i) 
< 2\pi $,
which respects the center U(1) symmetry
$\alpha_i \mapsto \alpha_i + {\rm const.}$.
\end{thebibliography}
\end{document}